\documentstyle[prc,preprint,aps,epsfig]{revtex}
\tighten
\begin{document}
\title{Inversion Potential Analysis of the
Nuclear Dynamics in the Triton}
\author{B.\ F.\ Gibson}
\address{Theoretical Division, Los Alamos National Laboratory\\
Los Alamos, New Mexico 87544\\ and \\
Institut f\"ur Kernphysik, Forschungszentrum J\"ulich\\
D--52425 J\"ulich, Germany}
\author{H.\ Kohlhoff and H.\ V.\ von Geramb}
\address{Theoretische Kernphysik, Universit\"at Hamburg\\
D--22761 Hamburg, Germany}
\author{G.\ L.\ Payne}
\address{Department of Physics \& Astronomy, University of Iowa\\
Iowa City, Iowa 52242}
\maketitle
\begin{abstract}
We report $^3$H binding energy calculations using inversion
potentials generated from phase shifts corresponding to
contemporary nucleon-nucleon potentials as well as modern
phase shift analyses.  We place limits upon local potential
triton binding energy calculations due to the underlying
uncertainties in their fit to the nucleon-nucleon phase
shifts: 7.7 $\pm$ 0.2 MeV.  We explore the role of the
nonlocality of momentum dependent potentials in the triton
binding energy.  In particular, we find the additional
binding energy in the case of the Bonn-B potential is due
to a long range nonlocality, which may correspond to a
three-nucleon force when the Hamiltonian is restricted to
local interactions.
\end{abstract}
\narrowtext
Triton model calculations, utilizing new Nijmegen local potentials
{\cite{stk93}} that fit the nucleon-nucleon scattering data in the
range of $0-350$ MeV almost as well as the Nijmegen phase shift
analysis, were recently reported {\cite{jlf93}}.  The results were
summarized as yielding a binding energy of 7.62 $\pm$ 0.01 MeV, some
0.86 MeV smaller than the experimental value of 8.48 MeV.
We examine here 1) the limits placed upon local potential
triton calculations by the uncertainties in the fits to the
nucleon-nucleon phase shifts, 2) the role of the nonlocality of
momentum dependent potentials in the triton binding energy, and 3) the
effects of long range nonlocality exhibited by certain contemporary
potential models.

Realistic potential models fitted to the nucleon-nucleon scattering
observables have been generated by several groups: for example,
Paris {\cite{paris}}, Nijmegen {\cite{nij78}}, and Bonn {\cite{bonnB}}.
The fits are at least semi-quantitative.  The limited number of
parameters in these potential models implies that the $\chi^2 / N$ fit
to the observables will necessarily be larger than that obtained in a
precision phase shift analysis.  By constructing partial-wave local
potentials, the Nijmegen group {\cite{stk93}} have succeeded in
obtaining
models  whose fits to the data are comparable to those obtained in
phase shift analyses.  Alternatively, the technique for constructing
inversion potentials from phase-shift data has been developed
to the point that one can generate equally precise partial-wave local
potentials  {\cite{kir89,hvg93,KTH93}}.  Thus, one can compare triton
binding energy results for potentials constructed from 1) a theoretical,
meson-exchange approach and 2) an inversion prescription that
generates an equivalent partial-wave local function.  Therefore,
the effect of nonlocality in the potential models can be investigated
quantitatively, both for short range nonlocality as one finds in the
Paris potential and for long range nonlocality as appears in the Bonn-B
potential.

There is no claim for a particular interaction model dynamics
in the inversion prescription.  The
physics resides in the assumption about the applicabilty
of a differential form of the Blankenbecler-Sugar
(relativistic Schr\"odinger) equation
with a local potential for each
partial wave. What we gain,
by dint of its construction with the Gelfand--Levitan--%
Marchenko integral equations, is that the resulting
partial-wave local potentials  reproduce the input phase shifts
along with the deuteron spectroscopic data; that is, the
physical observables and phase shifts
from which the potentials were generated
are exactly reproduced.  If nonlocality in the nucleon-nucleon
interaction can be shown to be mandated then, in the future,
it shall  be included in the inversion prescription.

The discrepancy between the experimental value for the triton
binding energy and results for various potential models has been
attributed by some to the need to include a three-body force
(3BF) in the Hamiltonian {\cite{3BF}}.  Carlson {\cite{carls}}
has shown that a phenomenological 3BF adjusted to reproduce the
triton binding energy will also yield a correct value for the
ground-state binding energy of the alpha particle.  Sauer and
collaborators {\cite{sauer}} have argued that the three-nucleon force,
which results when the $\Delta$ is eliminated from an $NN-N\Delta$
coupled-channel model of the nucleon-nucleon interaction,
contributes little to the triton binding.  Further model calculations
by Picklesimer and collaborators {\cite{pickl}} support this claim.
However, when the full Tucson-Melbourne (TM) three-nucleon
force {\cite{TM}} ($\pi \pi$, $\pi \rho$, and $\rho \rho$ terms
as recently published by Coon and Pe\~na {\cite{coon}}), which was
designed to be used with nucleon-nucleon potentials that incorporate
only nucleon-nucleon degrees of freedom, was combined
with the Reid soft-core {\cite{RSC}}, Paris {\cite{paris}}, and
Nijmegen {\cite{nij78}} potentials, Stadler {\it et al.}\
{\cite{sta93}} found that the model $^3$H binding energies
were close to 8.48 MeV.  Critics of this approach point out
that the meson-nucleon form factor cut-offs in the TM 3BF
are soft whereas those in the nucleon-nucleon potential models
are hard.  Nonetheless, the calculations demonstrate that a
3BF can play a role in the triton.  The source of that 3BF is
not necessarily established; it may come in part from nonlocality
in the nucleon-nucleon force.  Stadler {\it at al.}\
demonstrated that combining the TM 3BF with the Bonn-B potential
led to strong overbinding of the triton.  This implies
significant double counting when one combines the TM 3BF with
a nonlocal potential such as Bonn-B, unless the unknown
3BF associated with the Bonn-B potential proves to be
significantly larger than anticipated on the basis of
local potential 3BF model calculations.  As has been shown by
Polyzou and Gl\"ockle {\cite{poly}}, there is a unitary transform
relationship between Hamiltonians comprised of local NN plus NNN
potentials and Hamiltonians comprised of nonlocal NN potentials.
Their mathematical study did not enforce a requirement that the
interaction have a long range one-pion-exchange (OPE) tail,
but it is not difficult to conceive of restricting the
unitary transformations to ranges shorter than OPE.

Therefore, we wish to define the bounds on the triton
binding energy that result from the assumption of a local potential
derived directly from our current knowledge of the nucleon-nucleon
observables.  In addition, we will explore the consequences of the
long range nonlocality that exists in the Bonn-B potential model.
The source of this nonlocality within the model is not agreed upon,
but its role in the triton binding energy stands out clearly.

To establish the reliability of the inversion potential prescription,
we consider first the new Nijm-II model.  The $^2$H properties for this
potential as well as those of the Bonn-B and Paris momentum-dependent
potentials are summarized in Table \ref{tableone}.
Of the three potentials, only
the Nijm-II model contains explicit charge dependence.  While each
provides at least a semi-quantitative fit to the on-shell
nucleon-nucleon data below 300 MeV,
we are interested in comparing $^3$H binding
energy results for each model with those of the corresponding inversion
potential, obtained by using the pion subthreshold
model phases for $j \leq 2$ plus those potentials
from OPE for $2 < j \leq 4$ as input.

The $^2$H properties for the inversion potentials are listed in
Table \ref{tabletwo},
along with those resulting from the phase shift analysis
of Arndt {\cite{arndt}}.  The input for the inversion prescription,
in addition to the  model or experimental
phase shifts within  0--300  MeV,
include B$_2$, A$_s$, and $\eta$ respectively.
Above 300 MeV the phase shifts are smoothly extrapolated to
infinity with $\lim_{k\to\infty}\delta(k) \sim 1/ k$.
Comparing the remaining entries in the two tables, it should be
clear that the physical observables for the model deuterons are
all reasonably reproduced by the inversion potentials.
Even the model dependent D-state probability P$_D$ from the
Nijm-II and Paris potentials show quite similar entries in the
two tables.  However, we emphasize that P$_D$ for Bonn-B differs
noticeably between model and inversion potential.  That is, the
local inversion potential that is phase shift equivalent
to Bonn-B has a significantly larger P$_D$, one much closer to
that of Paris and Nijm-II.  It is the integrand
\begin{displaymath}
 P_D(r)={\int_r^\infty \Psi_D^2(x)\ dx \over \int_r^\infty
[ \Psi_S^2(x)+\Psi_D^2(x) ]\ dx},
\end{displaymath}
which demonstrates clearly that the difference in the deuteron wave
functions  arises at long range.  This is illustrated
in Figs. 1 and 2, for the Bonn-B and Paris potentials.
The difference in the case of Paris occurs essentially inside of
1 fm, whereas for the Bonn-B model differences persist out to
at least 3 fm.  In Ref{\cite{aga93}}, the tensor component of
the Bonn-B model was studied using a unitary transformation to
convert it to a local form; significant differences were found
well inside 2 fm which do not account for those we find
at longer range.  There is
built into the Bonn-B potential a long range nonlocality of unknown
origin.

How do the model $^3$H binding energies compare with the
$^3$H binding energies generated by their inversion potential
counterparts?  The trinucleon bound-state calculation was tested
using the Reid soft-core interaction.  Triton binding energies
for this partial-wave local potential, using a spline interpolation
on a mesh, agreed exactly with previously obtained {\cite{glp80}
model results.  Results for the Nijm-II model are listed in the
first line of Table \ref{tablethree}.  (For Nijm-II we have used
the effective charge-symmetric interaction {\cite{fgp87}} given by
$$
V(^1S_0) = \frac{2}{3} V_{nn} + \frac{1}{3} V_{np}
$$
to account for the charge dependence of the nuclear force.)
The difference between 7.62 MeV for the
Nijm-II model {\cite{jlf93}} and 7.60 MeV for its inversion potential
is attributed to   the different phase shifts between
300 MeV and infinity,   and   to the use of OPE for the higher
partial waves in the inversion potential calculation.  We
do  consider this as a measure of the agreement
(a null signal) in terms of the low energy uncertainties.
That is, starting from pion subthreshold phase shifts
coming from a partial-wave local potential,
we have generated a corresponding inversion potential which is
also partial-wave local and subthreshold phase equivalent, and
we have demonstrated that the two
yield the same $^3$H binding energy. The influence of
high energy phase shift variations  is small;
optimization of the phase shift extrapolation is
not a key issue.

We turn next to the Paris potential.  The fact that the $^2$H D-state
for the model and its inversion potential are very similar implies
that the momentum-dependent Paris potential is almost local.  This
is confirmed by the triton binding energies listed in
Table \ref{tablethree}. The
7.47 MeV from the model calculation {\cite{jlf88,bochm}} is close
to the 7.47 MeV from the inversion potential.
Comparing the results for the Paris potential with
those from the Paris inversion potential, one can see that there
is no significant nonlocality or phase shift extrapolation effect
noticeable for the Paris model.  The difference between
the Paris and Nijm-II results can be understood in part in terms of the
lack of charge dependence in the Paris $^1$S$_0$ potential model
{\cite{3BF,bochm}}.

The Bonn-B potential is considerably different, as we emphasized
above.  The $^2$H D-state probability difference between the
model and its local inversion potential signals this.  The triton
binding energies in Table \ref{tablethree} confirm it.
(We note that the difference in the value of B$_2$ in Tables I and II
is a measure of the numerical precision in this calculation; the
calculated numbers are of 5 digit accuracy at the 2-body level.)
The long range nonlocality
in the Bonn-B potential is a nontrivial aspect of the model.  If
one takes also into account the subthreshold model
phase shift differences such as those
due to the charge dependence omitted in the Bonn-B model,
then the result from the inversion Bonn--B
potential is quite close to that
for the Nijm-II model.   That is,
the $^3$H binding energy from the Bonn-B local inversion potential
is very close to that obtained from other local potentials which fit the
two nucleon data.  Therefore,
we conclude that phase differences between the Bonn-B and Nijm-II
models do not play a key role in determining the binding
energy of the triton.  In contrast, the remaining
difference between the 7.84
MeV from the inversion potential and the 8.14 MeV {\cite{bochm}} from
the original Bonn-B model is a clear measure of the effect that can
be produced by a long range nonlocality ($\geq$ 1 fm) in the
nucleon-nucleon interaction.

Finally we turn to the question of calculating the binding energy
of the triton using inversion potentials
for the Arndt phase shift analysis.
The  result is  listed in Table \ref{tablethree}
and the noticeable difference
compared to others is explained primarily by the
very  low energy phases.
The Arndt analysis was not given the same attention in this
region as was the Nijmegen analysis.  As a result, the $^3$H
binding energy for the Arndt inversion potential
does not agree with the inversion potential model results
discussed above.  In other words, the low energy phases (and Q)
of the Arndt analysis are inconsistent with the two-body
data, and the triton binding energy calculation confirms this.

Let us now consider the question of the uncertainty in the
triton binding energy due to uncertainties in the fits to the
underlying high energy nucleon-nucleon phase shifts.
The first element
of this analysis can be found in Table \ref{tablethree}.
The Nijm-II potential is said to
reproduce as precise a fit to the NN observables as does the
Nijmegen phase shift analysis for $0-300$ MeV laboratory incident
energies {\cite{jlf93}}.
However, the phases at higher energies are not strongly
constrained.  The Nijmegen group are careful to state that the phases
from their potentials are not to be taken as realistic outside of the
region in which they were fitted.  In fact, the Nijm-II phases are
somewhat pathological at energies higher than 300 MeV and their
potentials are thus not well behaved at short distances. We notice
also a short range attraction in the the $^1F_3,\ ^1P_1,\ ^3P_1$
and $^3F_3$ channels.  For that
reason we modified the Nijm-II phase shift extrapolation towards
the Arndt phase shifts {\cite{arndt}} at higher energies or took
simply a
smooth rational function extrapolation. This freedom in extrapolation
is difficult to eliminate and stresses as such the ill--posed aspect
of the problem and effective operator
nature of all  potentials.   Implementing an alternate choice of a
smooth extrapolation is straightforward  with inversion techniques.
Thus, the influence of the high energy phases upon
the $^3$H  binding energy can be  investigated without changing the low
energy phase shifts. Of great importance is the spectroscopic
deuteron bound state information; i.e., the binding energy and the
asymptotic normalization constants.  The
Nijmegen group extract these values from their potential and do not
fit these as independent data.
This input uncertainty  and its effect on the tensor potential
are  important.
The preferred optimal choice for the
Nijm-II inversion potential \cite{hvg93} yields the result for
the triton identified in Table \ref{tablethree} as Nijm-II$^*$.
The 70 keV difference between the two inversion potential triton binding
energies represents   a  minimum of  uncertainty
in local potential model calculations due to remaining phase shift
and  spectroscopic uncertainties.

A larger degree of uncertainty is implied by the triton results
summarized in Table \ref{tablefour}.
We used the Bonn-B inversion potential
for this study \cite{hvg93}}.
We smoothly
modified the $^1$S$_0$ phase shift function above 300 MeV such
that it  passed at
800 MeV  through  the  series of quoted
values.  The resulting inversion potentials produced the
triton binding energies listed in the
Table \ref{tablefour}.
Clearly, the high energy phase shifts do play
a  role in the triton binding energy. As expected however,
this is not as
large a difference as that coming from different models for
the low energy phase shifts and deuteron data.

Based upon these  studies of triton binding energy variations
due to limited constraints on higher energy phase shifts and the
Nijmegen potential model fits to the nucleon-nucleon observables, we
estimate that the $^3$H binding energy calculated using a
local potential model of the nucleon-nucleon
interaction should be
$$
B(^3{\rm H}) = 7.7 \pm 0.2 \; {\rm MeV}.
$$
The difference between this estimate and the experimental 8.48
MeV could be attributed to a three-body force contribution for
such a local potential {\it ansatz}.  An obvious question is
whether the Hamiltonian is better augmented by including
dynamically consistent three-body forces or by opening Pandora's
box to include nonlocal interactions.
The Bonn-B potential provides an
example of how longer range nonlocality can shift the triton
binding energy, although some significant fraction of this
may be attributed to the fact that the Bonn-B representation
of the $NN$ observables lies at the edge of the range of
accepted values rather than near the centroid.

\section*{Acknowledgement}
During the course of this work we had repeated communication with R.\
Machleidt and W.\ Gl\"ockle concerning use of the Bonn--B potential.
The work of the B.~F.~G.~was performed under the auspices of the
U.~S.~Department of Energy.  That of G.~L.~P.~was supported in part
by the U.~S.~Department of Energy.  B.~F.~G.~gratefully acknowledges
a Research Award for Senior Scientists by the Alexander von
Humboldt-Stiftung.  H.\ K.\ was supported in part by COSY,
KFA--J\"ulich, Grant Nr.\ 41126865.

\newpage
\begin{table}
\caption{$^2$H properties for the NN potential models.\label{tableone}}
\begin{tabular}{ccccccc}
Quantity & Nijm-II   &  Bonn-B &  Paris &   Experiment
\\  \hline
B$_2$ &  2.22458 & 2.22461 & 2.2249  & 2.22459 \cite{gre86} \\
P$_D$ &  5.66    &  4.99   & 5.77    &              \\
Q     & 0.2707   &  0.278  & 0.279   & 0.286\ \ \ \cite{bis79} \\
A$_s$ & 0.8847   &  0.8860 & 0.8866  & 0.88\ \ \ \ \cite{eri84} \\
$\eta$& 0.0252   &  0.0264 & 0.0261 &  0.026 \ \ \ \cite{rod86}\\
r$_{rms}$ & 1.9671 & 1.9688& 1.9716 & 1.96\ \ \ \ \cite{eri84}
\end{tabular}
\end{table}
\begin{table}
\caption{$^2$H properties for the inversion
potential models investigated.\label{tabletwo}}
\begin{tabular}{ccccc}
Quantity  &  Nijm-II &  Bonn-B &  Paris & Arndt   \\  \hline
B$_2$     & 2.2245(8) & 2.2246(5) & 2.2249(0) &  2.2246(0) \\
P$_D$     & 5.53    &    5.81 & 5.69   &  6.27    \\
Q         & 0.2705  &  0.2877 & 0.2788 &  0.2870  \\
A$_s$     & 0.8848  &  0.8861 & 0.8869 & 0.8860   \\
$\eta$    & 0.0252  &  0.0264 & 0.0261 & 0.0264   \\
r$_{rms}$ & 1.9672  &  1.9709 & 1.9716 & 1.9748
\end{tabular}
\end{table}
\begin{table}
\caption{Comparison of $^3$H binding energies in MeV for
model and inversion potentials.\label{tablethree}}
\begin{tabular}{ccc}
 Source &  BE(model) [MeV] &  BE(invers) [MeV]  \\ \hline
 Nijm-II &  7.62            &  7.60              \\
 Nijm-II$^*$ &              &  7.67              \\
 Paris  &  7.467           &  7.472             \\
 Bonn-B &  8.14            &  7.84              \\
 Arndt FA91 &              &  7.36
\end{tabular}
\end{table}
\begin{table}
\caption{Comparison of the Bonn inversion potential 34-channel
triton binding energy results in MeV for the $^1$S$_0$ phase shifts
fixed at 800 MeV at the values shown.\label{tablefour}}
\begin{tabular}{ccc}
 $\delta$ [degrees] & triton BE [MeV] &  comment \\  \hline
-35.62 &  7.87 &  \\
-41.15 &  7.85 &  \\
-43.04 &  7.84 &  $\sim$ Arndt\cite{arndt}\\
-44.28 &  7.83 &  \\
-45.01 &  7.82 &  \\
-47.97 &  7.77 &  \\
-54.93 &  7.62 &
\end{tabular}
\end{table}
\figure{Fig. 1: Comparison of the $^2$H wave function components from
the Bonn-B model and its inversion potential counterpart along with
the integral for P$_D$ (scaled by 10).
Solid line corresponds to the inversion
potential, dashed line to the original model.
\label{figureone}}
\figure{Fig. 2: Comparison of the $^2$H wave function components from
the Paris model and its inversion potential counterpart along with
the integral for P$_D$ (scaled by 10).
Solid line corresponds to the inversion
potential, dashed line to the original model.
\label{figuretwo}}

\newpage postscript figures \\[3cm]
\unitlength1.0cm
\begin{figure}\centering
\begin{picture}(15,8)(0.0,1.0)\centering
\epsfig{figure=bonrwave.eps,width=\textwidth}
\end{picture}
\end{figure}
\newpage postcript figures \\[3cm]
\unitlength1.0cm
\begin{figure}\centering
\begin{picture}(15,8)(0.0,1.0)\centering
\epsfig{figure=parrwave.eps,width=\textwidth}
\end{picture}
\end{figure}
\end{document}